\documentclass[11pt]{article}
\usepackage{graphicx,amsmath,amsthm,mathrsfs,amssymb,float}
\usepackage[usenames]{color}
\usepackage{ulem}

\newcommand{\add}[1]{{\color[rgb]{1,0,0} #1}}

\newcommand{\ee}{\end{equation}}
\newcommand{\word}[1]{\,\,\mbox{#1}\,\,}
\newcommand{\reff}[1]{(\ref{#1})}
\newcommand{\beq}{\begin{equation}}
\newcommand{\eeq}[1]{\label{#1}\end{equation}}
\newcommand{\beg}{\begin{equation*}}
\newcommand{\eeg}{\end{equation*}}

\newcommand{\eq}{\!=\!}
\newcommand{\p}{\!+\!}
\newcommand{\m}{\!-\!}

\newcommand{\bsplit}{\begin{split}}
\newcommand{\esplit}{\end{split}}

\begin{document}
\def\theequation{\arabic{section}.\arabic{equation}}
\begin{titlepage}
\title{Radiatively induced symmetry breaking and the conformally coupled
magnetic monopole in AdS space}
\author{$^{1,2}$Ariel Edery\thanks{Email: aedery@ubishops.ca}\,\,,\quad $^{3}$Noah Graham\thanks{Email:
ngraham@middlebury.edu}\\\\$^1$ {\small\it Department of Physics, Bishop's University,
2600 College Street, Sherbrooke, QC
J1M~1Z7}\\$^2${\small\it Kavli Institute for Theoretical Physics, University of California,}
\\{\small\it Kohn Hall, Santa Barbara, CA 93106 U.S.A.}\vspace{0.5em} \\ $^3${\small\it Department of
Physics, Middlebury College, Middlebury, VT 05753} \vspace{0.5em}
\\
}
\date{} \maketitle
\vspace*{-1truecm}
\begin{abstract}
We implement quantum corrections for a magnetic monopole in a classically conformally invariant theory containing gravity. This yields the trace (conformal) anomaly and introduces a length scale in a natural fashion via the process of renormalization. We evaluate the one-loop effective potential and extract the vacuum expectation value (VEV) from it; spontaneous symmetry breaking is radiatively induced. The VEV is set at the renormalization scale $M$ and we exchange the dimensionless scalar coupling constant for the dimensionful VEV via dimensional transmutation. The asymptotic (background) spacetime is anti-de Sitter (AdS) and its Ricci scalar is determined entirely by the VEV. We obtain analytical asymptotic solutions to the coupled set of equations governing gravitational, gauge and scalar fields that yield the magnetic
monopole in an AdS spacetime.
\end{abstract}
\setcounter{page}{1}
\end{titlepage}

\def\theequation{\arabic{section}.\arabic{equation}}


\section{Introduction}

The t'Hooft Polyakov magnetic monopole \cite{tHooft, Polyakov} is a
toplogical soliton solution of non-Abelian gauge theory. When the
gauge fields are coupled to scalar fields with the typical symmetry
breaking potential $V(\phi)=\tfrac{\lambda}{4}(\phi^4-a^2\,\phi^2)$,
the $SU(2)$ gauge symmetry gets broken down to the residual $U(1)$
gauge of the electromgnetic field leading to a magnetic monopole. The
spontaneous breaking of gauge symmetry leaves the 10 parameter
Poincar\'{e} group intact. A few years ago, it was shown that one can
add the gravitational field to this picture and enlarge the symmetry
group to the 15 parameter conformal group SO(2,4) by conformally
coupling the scalar fields to the metric \cite{MB}. The original gravity sector does not
include an Einstein-Hilbert term because that term
is not conformally invariant; instead, one uses the
conformally invariant Weyl squared term. A
cosmological constant is also forbidden by conformal
invariance. In this massless
model, the usual $a^2 \phi^2$ term is replaced by
the conformal coupling term $-R\,\phi^2/6$,
where $R$ is the Ricci scalar. The vacuum expectation value (VEV) is now expressed in terms of $R$. The theory is classically scale invariant,
and one must therefore choose a VEV value by hand (or alternatively specify the asymptotic value of the Ricci scalar). Magnetic monopole solutions in a background anti-de Sitter
(AdS) space were then found numerically \cite{MB}. The static
sherically symmetric solution at large distances (outside the monopole
core) is Schwarzschild-AdS. It should be stressed that the solution in \cite{MB} is not
a black hole (BH) solution since the core is non-singular
and static. By definition, a BH has a horizon and there exits a region in the
interior which is non-stationary, where all Killing vectors are
spacelike \cite{Edery2} (extremal BHs are the exception since their
interiors can be static or stationary, i.e. they
possess a timelike Killing vector \cite{Edery2}).

In this work, we introduce a length scale
by considering the effects of quantum fluctuations of scalar fields at
one loop in a classical gravitational and gauge field background. 
We calculate the contribution of
the trace anomaly through the Schwinger-Dewitt
coefficient $a_2(x)$ of the theory. To determine the vacuum
expectation value (VEV), we derive the effective potential at one loop
for a massless theory with a triplet of scalar fields and $R \phi^2/6$
and $\lambda \phi^4/4!$ interactions. The coupling constant $\lambda$
is a running coupling constant which is defined at a renormalization
scale $M$. We use this fact to trade the dimensionless $\lambda$ for
the dimensionful VEV of the theory, through dimensional transmutation \cite{Coleman}. The asymptotic (background) spacetime is
AdS and its Ricci scalar is completely determined by the VEV. The effective
potential is also used to evaluate the composite operator
$[E]$ that enters the quantum-corrected equations of motion. Finally, we derive the equations of
motion governing the gravitational, gauge and scalar fields and solve
them analytically in the asymptotic region to obtain the relation between the Ricci scalar and the VEV.

Though conformal invariance is presented here in the context of the
magnetic monopole, there is a more general interest in scale and
conformal symmetry as a possible fundamental principle in physics and
cosmology, as discussed for example in
\cite{Bars}. As described there, the classical action of the
standard model is already consistent with global scale symmetry if the
Higgs mass is dropped. Furthermore,
on cosmic scales, the nearly scale-invariant spectrum of primordial
fluctuations seems to demand an explanation based on a fundamental
symmetry in nature. In short, the authors argue that scale and
conformal symmetry may be the clue to fitting observations on very
small and very large scales in a coherent theory (see \cite{Bars} for
more details and references). The conformal anomaly has also attracted interest 
as an approach to the cosmological constant problem \cite{Thomas}. The argument put forward is that 
the cosmological constant problem may arise from the infrared sector of the effective theory of gravity 
where the conformal anomaly is expected to be relevant.

\section{Quantum corrections}

Quantum vacuum fluctuations break explicitly the global scale symmetry and hence the entire conformal symmetry. The energy-momentum tensor develops a nonzero trace that can be evaluated via the
Schwinger-Dewitt coefficient $a_2(x)$ in four dimensions. This is the
well known trace or conformal anomaly \cite{Mukhanov,Birrell}. The effects due
to vacuum fluctuations of matter fields can be included via an
effective action $W$. It is convenient to split $W$ into two parts: a
divergent part $W_{\rm div}$ and a renormalized finite part $W_{\rm ren}\eq W
\m W_{\rm div}$. $W_{\rm div}$ is due to high-frequency fluctuations and is
therefore local and state-independent \cite{Mukhanov}. It is expressed
as an integral over $a_2(x)$, which can be calculated via a set of
``curvatures" \cite{Barvinsky}. $W_{\rm div}$ is incorporated into the original action $S$ by adding
the appropriate counterterms. We label this $S_{ren}$, the renormalized local part of the one loop effective action. $W_{\rm ren}$ is the finite state-dependent nonlocal part of the effective action due to long wavelength
fluctuations that sample the entire geometry. For example, the finite part of the vacuum expectation value of the energy-momentum tensor, $\langle T_{\mu\nu}\rangle$, that will appear on the right hand side of the gravitational equations of motion, can be viewed as a Casimir effect which also arises from long wavelength quantum fluctuations \cite{Mostepanenko}. 

\subsection{Schwinger-Dewitt coefficient for the monopole action and trace anomaly}       
   
The conformally invariant action that leads to a magnetic monopole coupled to gravity contains a metric $g_{\mu\nu}$, a triplet of scalar fields $\phi^{\,a}$ and nonabelian gauge fields $A_{\mu}^{\,a}$. It is given by \cite{MB} 
\begin{align}
S&=\int d^4x\,\sqrt{-g}\,\Big(C_{\mu\nu\sigma\tau}C^{\mu\nu\sigma\tau}-\dfrac{1}{4e^2}\,F_{\mu\nu}^{\,a}\,F^{\mu\nu}_a + D_{\mu}\phi^a\,D^{\mu}\phi_a +\dfrac{1}{6}\,R\,\phi^a\phi_a - \dfrac{\lambda}{4!}(\phi^a\,\phi_a)^2 \Big)\nonumber\\
&=\int d^4x\,\sqrt{-g}\,\Big(C^2-\dfrac{1}{4e^2}\,F^2 + (D\phi)^2 +\dfrac{1}{6}\,R\,\phi^2 - \dfrac{\lambda}{4!}\phi^4 \Big)
\label{action2}\end{align}
where $C_{\mu\nu\sigma\tau}$ is the Weyl tensor, $F_{\mu\nu}^{\,a}$ is
the gauge field strength defined by  $\nabla_{\mu}\,A_{\nu}^{\,a}\m
\nabla_{\nu}\,A_{\mu}^{\,a} \p
\varepsilon^{a}_{\,\,\,bc}\,A_{\mu}^{\,b} \,A_{\nu}^{\,c}\,$, and
$D_{\mu}\phi^{\,a}$ is the covariant derivative defined by
$\nabla_{\mu}\,\phi^{\,a} \p \varepsilon^{a}_{\,\,\,bc}\,A_{\mu}^{\,b}
\,\phi^{\,c}$. The square of the Weyl tensor is denoted as $C^2$ and
$F^2\equiv F_{\mu\nu}^{\,a}\,F^{\mu\nu}_a$, $(D\phi)^2\equiv
D_{\mu}\phi^a\,D^{\mu}\phi_a$ and $\phi^2\equiv\phi^a\phi_a$. The
Ricci-Levi-Civita three-dimensional coefficients $\varepsilon_{abc}$
are totally antisymmetric. Raising or lowering of
the internal (latin) indices does not change the sign so that
$\phi_a=\phi^a$ and implicit summation is assumed throughout
e.g. $\phi_a\phi^a\eq\phi_1^{\,2}+\phi_2^{\,2}+\phi_3^{\,2}$. The action \reff{action2} is invariant under the conformal transformations $g_{\mu\nu}\to \Omega^2(x) g_{\mu\nu}$ and $\phi^a \to \phi^a/\Omega(x)$ where $\Omega(x)$ is called the conformal factor, an arbitrary positive smooth function.  

We consider the vacuum fluctuations of the triplet $\phi^{\,a}$ in a
gravitational and gauge field classical background. The Euclidean
effective action is given by $W= \tfrac{1}{2} \,\text{tr} \,(
\,\text{ln}\,\hat{H}\,)$ where $\hat{H}$ is the hessian of the
Euclidean version of the action \reff{action2}.
It takes the following form \cite{Barvinsky} 
\beq
\hat{H}= g^{\mu\nu}D_{\mu}D_{\nu} + \hat{P} - \dfrac{1}{6}R \hat{1}\,.
\eeq{hessian}
The above operator is a $3\times3$ matrix in the internal vector space
of the triplet $\phi^a$ and $\hat{1}$ is the identity matrix. Here $\hat{P}$ arises from the
self interaction term $\lambda(\phi^a\,\phi_a)^2/4!$, $\tfrac{1}{6}R \hat{1}
$ arises from the conformally
coupled scalar field term $\tfrac{1}{6}\,R\,\phi^a\phi_a$ and
$g^{\mu\nu}D_{\mu}D_{\nu}$ arises from the covariant derivative term
$D_{\mu}\phi^a\,D^{\mu}\phi_a$. The gravitational and nonabelian gauge
fields are treated classically. One identifies three ``curvatures''
for the operator \reff{hessian} that enter into the calculation of the
effective action \cite{Barvinsky}.  These are the Riemann curvature
associated with $g_{\mu\nu}$, the commutator curvature
$\hat{\mathscr{R_{\mu\nu}}}$ associated with the covariant derivative
$D_{\mu}$ and defined by
\beq
[D_{\mu},D_{\nu}]\,\phi^{\,a}=\mathscr{R}^{\,a}_{\;\;\;b \,\mu\,\nu}\,\phi^{\,b} \word{with} \hat{\mathscr{R_{\mu\nu}}}\equiv \mathscr{R}^{\,a}_{\;\;\;b \,\mu\,\nu} \eeq{commutator} and the potential $\hat{P}$ which is its own curvature.
The divergent part $W_{\rm div}$ of the Euclidean effective action can be expressed entirely in terms of these curvatures as \cite{Barvinsky}
\begin{align}\label{Wdiv}
W_{\rm div}&= \dfrac{1}{n-4} \int d^{\,4}x \,\sqrt{g} \;\;\text{tr}\,\hat{a}_2(x,x) \quad\quad (n\to 4)\\
\hat{a}_2(x,x)&=\dfrac{1}{180}\Big(R_{\mu\nu\sigma\tau}R^{\mu\nu\sigma\tau} \m R_{\mu\nu}R^{\mu\nu} \p \,\square \,R\Big)\hat{1}
\p \dfrac{1}{12}\,\hat{\mathscr{R_{\mu\nu}}}\hat{\mathscr{R^{\mu\nu}}} \p \dfrac{1}{2}\,\hat{P}^2 \p \dfrac{1}{6}\,\square \,\hat{P}\,.
\end{align}
We now evaluate the curvatures $\hat{P}$ and $\hat{\mathscr{R_{\mu\nu}}}$ for the Euclidean version of the action \reff{action2}. The matrix elements of the potential $\hat{P}$ are 
\begin{align}
P_{ij}&= -\dfrac{\partial}{\partial\phi_i} \dfrac{\partial}{\partial\phi_j} \dfrac{\lambda}{4!} \big(\phi_a\phi^a\big)^2=-\dfrac{\lambda}{6}\,(\delta_{ij}\,\phi_a\phi^a+ 2\,\phi_i\,\phi_j\,)\word{where} i,j=1,2,3 \,.
 \label{pij}\end{align}
To obtain $\hat{\mathscr{R_{\mu\nu}}}$, we first evaluate the double covariant derivative, 
\begin{align}
\!\!\!\!\!D_{\mu}D_{\nu}\phi^{\,a}&= \nabla_{\mu}(D_{\nu}\phi^{\,a}) + \varepsilon^{a}_{\,\,\,bc}\,A_{\mu}^{\,b} \,D_{\nu}\phi^{\,c}\nonumber\\
&=\nabla_{\mu}(\nabla_{\nu}\phi^{\,a} + \varepsilon^{a}_{\,\,\,de}\,A_{\nu}^{\,d} \,\phi^{\,e})+ \varepsilon^{a}_{\,\,\,bc}\,A_{\mu}^{\,b} (\nabla_{\nu}\phi^{\,c} +\varepsilon^{c}_{\,\,fg}\,A_{\nu}^{\,f} \,\phi^{\,g})\nonumber\\&=\nabla_{\mu}\nabla_{\nu}\phi^a \p \varepsilon^{a}_{\,\,\,de}\,\nabla_{\mu}A_{\nu}^{\,d}\,\phi^{\,e}\p \varepsilon^{a}_{\,\,\,de}\,A_{\nu}^{\,d}\nabla_{\mu}\phi^{\,e} \p \varepsilon^{a}_{\,\,\,bc}\,A_{\mu}^{\,b} \nabla_{\nu}\phi^{\,c} \p \varepsilon^{a}_{\,\,\,bc}\,\varepsilon^{c}_{\,\,fg}\,A_{\mu}^b\,A_{\nu}^{\,f} \,\phi^{\,g}\,.\end{align}
The commutator is then given by       
  \begin{align}
[D_{\mu},D_{\nu}]\phi^{\,a} &= \varepsilon^{a}_{\,\,\,de}\,\big(\nabla_{\mu}A_{\nu}^{\,d}-\nabla_{\nu}A_{\mu}^{\,d}\big)\,\phi^{\,e} + \varepsilon^{a}_{\,\,\,bc}\,\varepsilon^{c}_{\,\,fg}\big(A_{\mu}^b\,A_{\nu}^{\,f} -A_{\nu}^b\,A_{\mu}^{\,f}\big)\, \phi^{\,g}\nonumber\\\nonumber&=
\varepsilon^{a}_{\,\,\,de}\,\big(\nabla_{\mu}A_{\nu}^{\,d}-\nabla_{\nu}A_{\mu}^{\,d} + \varepsilon^{d}_{\,\,fg}\,A_{\mu}^f\,A_{\nu}^{\,g}\big) \phi^{\,e}\\&= \varepsilon^{a}_{\,\,\,de}\,F_{\mu\nu}^{\,d}\,\phi^e\label{commutator2}\end{align}
where the contracted epsilon identity $\varepsilon_{jk\ell}\,\varepsilon_{jmn}\eq\delta_{km}\delta_{\ell\,n}\m\delta_{kn}\delta_{\ell m}$ was used. Comparing \reff{commutator2} with \reff{commutator}, we obtain the commutator curvature \beq \hat{\mathscr{R_{\mu\nu}}}\equiv\mathscr{R}^{\,a}_{\;\;\;e \,\mu\,\nu} \eq\varepsilon^{a}_{\,\,\,de}\,F_{\mu\nu}^{\,d}\,.\eeq{commute}
The trace of the operators that appear in $\hat{a}_2(x,x)$, evaluated using the formulas \reff{pij} and \reff{commute} for $\hat{P}$ and $\hat{\mathscr{R_{\mu\nu}}}$ respectively, are 
\beq
\text{tr}\,\hat{1} =3\;,\; \text{tr}\,\hat{\mathscr{R_{\mu\nu}}}\hat{\mathscr{R^{\mu\nu}}}=-2\,F^2\;,\;
\text{tr} \,\hat{P}^2=\dfrac{22}{3}\dfrac{\lambda^2}{4!}\,\phi^4\;,\;\text{tr}\,\square\,\hat{P}= -\dfrac{5}{6} \,\lambda\,\square\,\phi^2\,.
\eeq{dse}
The quantity of interest that appears in the integrand of $W_{\rm div}$ is  
\begin{align}\label{a2}
a_2(x)\equiv\text{tr}\,\hat{a}_2(x,x)=\dfrac{1}{60}\Big(R_{\mu\nu\sigma\tau}R^{\mu\nu\sigma\tau} &\m R_{\mu\nu}R^{\mu\nu} \p \,\square \,R\Big)\\\nonumber
&-\dfrac{1}{6}\, F^2 + \Big(\dfrac{11}{3}\Big)\dfrac{\lambda^2}{4!}\phi^4 - \dfrac{5}{36}\,\lambda \square \,\phi^2 \end{align}
where the factor of $11/3$ in front of $\lambda^2$ stems
from having a triplet of scalar fields (it is equal to
$3$ for a single scalar field). To cancel the divergent part
$W_{\rm div}$, we simply add the appropriate counterterm to the original
action \reff{action2}. The renormalized local part of the action at one loop is simply
\beq
S_{\rm ren}=\int d^4x\,\sqrt{-g}\,\Big(\alpha \,R^2 +\beta
\,R_{\mu\nu}R^{\mu\nu} -\dfrac{1}{4e^2}\,F^2 + (D\phi)^2
+\dfrac{1}{6}\,R\,\phi^2 - \dfrac{\lambda}{4!}\,\phi^4 \Big)\,.
\eeq{action3} 
The total derivatives $\square R$ and $\square \,(\phi_a\,\phi^a)$ appearing in \reff{a2} are not included in the above action as they lead to boundary terms that have no bearing on the equations of motion. The
terms $C_{\mu\nu\sigma\tau}C^{\mu\nu\sigma\tau}$ and
$R_{\mu\nu\sigma\tau}R^{\mu\nu\sigma\tau}$ do not appear in
\reff{action3} since they can be eliminated in favor of
$R_{\mu\nu}R^{\mu\nu}$ and $R^2$ using the equality
$C_{\mu\nu\sigma\tau}C^{\mu\nu\sigma\tau}=
R_{\mu\nu\sigma\tau}R^{\mu\nu\sigma\tau} \m 2\,R_{\mu\nu}R^{\mu\nu}
+R^2/3$ and the fact that the integral of
$R_{\mu\nu\sigma\tau}R^{\mu\nu\sigma\tau} \m 4\,R_{\mu\nu}R^{\mu\nu}
\p R^2$, the Gauss-Bonnet integral, is a topological invariant (the
Euler number) in four dimensions. The quantities $e$,
$\lambda$, $\alpha$, and $\beta$ are
renormalized constants. They are running coupling constants governed by a
renormalization group equation which we do not need to state here for
our purposes (for details see \cite{CollinsBrown,Hathrell}). At one
loop, the constants $\alpha$ and $\beta$ are
related by $\alpha=-\beta/3$ and the
sum of the first two terms in \reff{action3} are
conformally invariant. At higher loops, non-conformally invariant
terms like $R^2$ are generated \cite{CollinsBrown,Hathrell} and the
two constants become independent. Our calculations will be at one loop, so we assume 
the relation $\alpha=-\beta/3$.

\subsection{Anomalous trace}

To find the trace, we first note that under a conformal tranformation the inverse metric transforms according to $g^{\mu\nu}(x)\to e^{2\sigma(x)} g^{\mu\nu}(x)$ and the scalar field transforms according to $\phi^a(x) \to e^{\sigma(x)} \phi^a(x)$ where $\sigma(x)$ is introduced for convenience and is related to the conformal factor $\Omega(x)$ (previously introduced) by $e^{-\sigma(x)}\equiv \Omega(x)$. This yields the functional relation
\beq
\dfrac{\delta}{\delta \sigma(x)}= 2 g^{\mu\nu}(x)\dfrac{\delta}{\delta g^{\mu\nu}(x)} + \phi^a(x)\dfrac{\delta}{\delta \phi^a(x)} \,.
\eeq{functional}
The functional variation of the one-loop effective action with respect to $\sigma(x)$ is proportional to the Scwinger-Dewitt coefficient $a_2(x)$ \cite{Barvinsky} 
\beq
\dfrac{\delta\,W_{\rm ren}}{\delta \sigma}= \dfrac{\sqrt{-g}\,a_2(x)}{16 \,\pi^2}
\eeq{deltaW}
and using \reff{functional} this yields 
\beq
\dfrac{1}{\sqrt{-g}}\big(2\,g^{\mu\nu}\dfrac{\delta W_{\rm ren}}{\delta g^{\mu\nu}}+\phi^a \dfrac{\delta W_{\rm ren}}{\delta \phi^a}\big)=\dfrac{a_2(x)}{16 \pi^2}\,.
\eeq{deltaW2}
The vacuum expectation value of the energy-momentum tensor is
defined as
\beq
\langle T_{\mu\nu}\rangle=\dfrac{2}{\sqrt{-g}}\dfrac{\delta W_{\rm ren}}{\delta g^{\mu\nu}}\,.
\eeq{EMT1}
From \reff{deltaW2}, its trace is given by   
\beq
\langle T^{\mu}_{\mu}\rangle = \dfrac{a_2(x)}{16 \pi^2}- [E]
\eeq{trace2}
where 
\beq
[E] \equiv \dfrac{\phi^a}{\sqrt{-g}} \dfrac{\delta W_{\rm ren}}{\delta \phi^a} 
\eeq{E} 
and 
\beq
a_2(x)=\dfrac{1}{60}\Big(R_{\mu\nu\sigma\tau}R^{\mu\nu\sigma\tau} \m R_{\mu\nu}R^{\mu\nu} \p \,\square \,R\Big)
-\dfrac{1}{6}\, F^2 + \dfrac{11}{3} \dfrac{\lambda^2}{4!} \phi^4 - \dfrac{5}{36}\,\lambda \square \,[\phi^2]\,.
\eeq{aa2}
The renormalized composite operator $[E]$ can be obtained readily once the effective potential has been calculated.

\subsubsection{AdS space}

As already discussed, the magnetic monopole solution is obtained when the spacetime is asymptotically anti-de Sitter space. This is a maximally symmetric spacetime with constant Ricci scalar $R$ (positive in our notation). It is the submanifold obtained by embedding a hyperboloid in a flat five-dimensional spacetime of signature(+,+,-,-,-). The universal covering space of AdS space has the topology of $R^4$ and can be represented by the following metric:
\beq
ds^2= (1+k\,r^2)\,dt^2 -\dfrac{dr^2}{1+k\,r^2} -r^2\big(d\theta^2+\sin^2\theta\,d\phi^2\big)
\eeq{AdS}   
where $(r,\theta, \phi)$ cover the usual range of spherical coordinates ($r\!\ge\!0\,$, $0\le\!\theta\le\!\pi\,$, $0\le\!\phi\!<2\,\pi$) and 
$-\infty<\!t<\!\infty$. The constant $k$ is positive with the Ricci
scalar given by $R=12k$. 

The symmetry group (isometry group) of AdS space is the ten parameter group SO(2,3). The only maximally form invariant rank two tensor under this group is the metric $g_{\mu\nu}$ (times a constant) so that the expectation value of the energy-momentum tensor in AdS space can be expressed in terms of its trace: 
\beq
\langle T_{\mu\nu}\rangle_0 = \dfrac{1}{4} \,g_{\mu\nu}\langle
T^{\mu}_{\,\,\mu}\rangle_0
\eeq{TT}  
where the zero subscript means that the quantity is evaluated asymptotically in AdS space, the vacuum spacetime.

\section{The effective potential for a massless theory with $R\phi^2$ and $\lambda \phi^4$ interactions}

In this section we obtain the one loop effective potential by summing
all the one loop one-particle irreducible (1PI) Feynman diagrams in
the presence of $\lambda \phi^4/24$ and $-R \phi^2/6$
interactions. The coupling constant $\lambda$ is defined at a
renormalization scale $M$. We choose the renormalization scale $M$
to coincide with the VEV, the
minimum of the effective potential. This in turn fixes the value of
$\lambda$; the dimensionless constant $\lambda$ is traded for the dimensionful VEV through dimensional transmutation. The expectation value of the composite operator $[E]$, which appears both in the trace anomaly and in the equations of motion for the scalar field, can be readily obtained from the effective potential. Note that at one loop the effective potential and the expectation value of composite operators do not generate new geometrical curvature terms; they are generated starting only at two loops \cite{CollinsBrown, Hathrell}. In particular, for the calculation of the effective potential, the $\sqrt{-g}$ factor plays no role at one loop. The calculation proceeds in the same manner as in flat space, though of course the Ricci scalar $R$ is non-zero and acts as a vertex for the $R\phi^2$ interaction. 

\begin{figure}
\begin{center}
\includegraphics[scale=1]{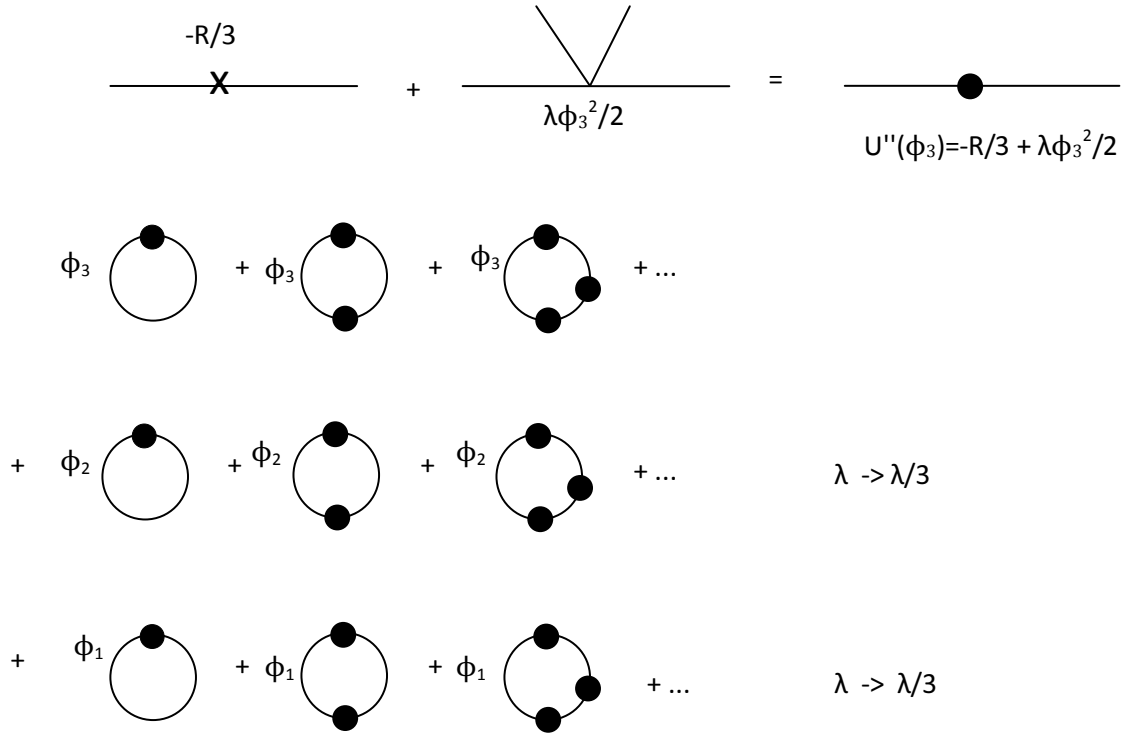}
\caption{One loop Feynman diagrams for a triplet of scalar fields.
  The two vertices corresponding to $-R\phi^2/6$ and $\lambda
  \phi^4/24$ interactions, which are replaced by a single vertex (black dot). The last two sets of diagrams, where the components $\phi_2$ and $\phi_1$ run around the loop, have a vertex with $1/3$ the coupling constant $\lambda$ of the first set of diagrams. The propagator is massless.}
\label{RPhi2}
\end{center}
\end{figure}

The (classical) potential is given by 
\beq
U= \dfrac{\lambda}{4!} \,\phi^4 -\dfrac{1}{6} R \,\phi^2 \,.
\eeq{Vcl}
We have a triplet of scalar fields and without loss of generality we
take the vacuum expectation value to lie along the third component
$\phi_3$. For $\phi_3$ loops, $U$ generates two vertices: $-R/3$ and $\lambda
\phi_3^2/2$. They can be combined into a single vertex given by the
second derivative $U''(\phi_3)=-R/3 +\lambda \phi_3^2/2$. The vacuum
expectation value of $\phi_1$ and $\phi_2$ are zero but they can still
fluctuate in loops. The vertex for both is
$U''(\phi_1)= U''(\phi_2)= -R/3 + \lambda \phi_3^2/6$ which is
equivalent to replacing the coupling constant $\lambda$ by $\lambda/3$
in the previous case. There are three sets of one loop 1PI Feynman
diagrams; these are depicted in Fig.\ref{RPhi2}. Let the classical field $\phi_{c}(x)$ be defined as the vacuum expectation value of $\phi_3$ in the presence of some external source $J(x)$ 
\beq
\phi_{c}(x)=\left.\dfrac{\langle 0|\phi_3(x)|0\rangle}{\langle 0|0\rangle}\right|_J.
\eeq{Phi0}  
where $J$ appears in the action in the usual fashion via the source
term $J \phi$. The effective potential is obtained by summing all the
diagrams in Fig.\ref{RPhi2}. Note that the propagator is massless. For the first set of diagrams, the one-loop contribution yields 
\beq
\begin{split}
V&=i\int\dfrac{d^4k}{(2\pi)^4} \sum_{n=1}^{\infty} \dfrac{1}{2n}\Bigg[\dfrac{-R/3 +\lambda \phi_{c}^2/2}{k^2+i\epsilon}\Bigg]^{n}\\
&= \dfrac{1}{16\pi^2} \int_0^{\Lambda} k_E^3  \ln\Bigg[1+ \dfrac{-R/3 +\lambda \phi_{c}^2/2}{k_E^2}\Bigg]
\end{split}
\eeq{V}
where $k_E$ is the Euclidean momenta and $\Lambda$ is a momentum cut-off. The integral in \reff{V} can be readily evaluated but we do not write it out explicitly here. The other two sets of Feynman diagrams can be evaluated by simply replacing $\lambda$ by $\lambda/3$ in \reff{V}. The one loop contribution to the potential is then
\beq
V_1= V + 2\,V[\lambda\rightarrow\lambda/3]\,.
\eeq{V1}
As it stands, the expression $V_1$ is divergent in the infinite $\Lambda$ limit. This is handled in the usual fashion by adding the necessary counterterms and then imposing the appropriate renormalization conditions. The total potential is given by
\beq
V_{tot}=\dfrac{\lambda}{4!} \,\phi_c^4 -\dfrac{1}{6} R \,\phi_c^2 + V_1 + A \phi_{c}^2 + B \phi_{c}^4
\eeq{V2}
where the last two terms are the counterterms. The constants $A$ and $B$ are determined via the renormalization conditions
\beq
\lambda=\left.\dfrac{d^4V_{tot}}{d\phi_{c}^4} \right|_{\phi_c=M} \word{and} -\dfrac{R}{3}= \left.\dfrac{d^2V_{tot}}{d\phi_{c}^2} \right|_{\phi_c=0}\,.
\eeq{Renorm}
The renormalization scale $M$ sets the scale for the theory. Substituting $A$ and $B$ back into \reff{V2}, taking the infinite $\Lambda$ limit and then collecting terms into compact expressions, we obtain the one loop effective potential 
\beq
\begin{split}
V_{\rm eff}=  \dfrac{\lambda}{4!} \,\phi_c^4 -\dfrac{1}{6} R \,\phi_c^2 + \dfrac{1}{1152 \pi^2}&\Bigg[18\,\Big(\!-\dfrac{R}{3} + \dfrac{\lambda \phi_c^2}{2}\Big)^2 
  \ln\Big[\,\dfrac{2 R - 3 \lambda \phi_c^2}{2 R}\,\Big]\\&\quad+ 36\,\Big(\!-\dfrac{R}{3} + \dfrac{\lambda \phi_c^2}{6}\Big)^2 
  \ln\Big[\,\dfrac{2 R - \lambda \phi_c^2}{2 R}\,\Big]+ 5 R \,\lambda\, \phi_c^2 \\&\qquad+ \lambda^2 \phi_c^4 \Big(\dfrac{9}{2}\,\ln\Big[\,\dfrac{2 R}
  {2 R- 3 \lambda M^2}\,\Big]
 +\ln\Big[\,\dfrac{2 R}
  {2 R- \lambda M^2}\,\Big]\Big)\Bigg]\\ &\qquad\quad+ \Delta
\end{split}
\eeq{Veff} 
where
\beq
\Delta=\lambda^2 \phi_c^4\,\dfrac{(-1584 R^4+ 11904 M^2 R^3 \lambda - 20360 M^4 R^2 \lambda^2 + 
    12480 M^6 R \lambda^3 - 
    2475 M^8 \lambda^4)}{13824 \,\pi^2(-2 R + 
    M^2 \lambda)^2 (-2 R + 3 M^2 \lambda)^2}\,.
\eeq{Chi}
Let $\phi_c=v$ be the vacuum expectation value (VEV). It takes on this
value in the asymptotic (background) spacetime, which is AdS space. We
will see later, in section 4.1, that solving the equations of motion asymptotically yields the
relation \reff{RAdS} between $R_0$, the Ricci scalar of AdS space, and the VEV:
$R_0=\sqrt{110}\, \lambda \,v^2$. The VEV occurs at the minimum of
the effective potential in the AdS background spacetime, where
\beq
\dfrac{dV_{\rm eff}}{d\phi_c}\bigg|_{\substack{\phi_c=v\\R=\sqrt{110}\lambda\,v^2}}=0\,.
\eeq{minimum}
We set the arbitrary scale $M$
to be equal to the VEV
i.e. $\phi_c\eq v \eq M$. Equation
\reff{minimum} then yields a numerical value of $\lambda=C/D=2519.926$,
where the exact expressions for $C$ and $D$ are 
\beq
\begin{split}
C&=288 \pi^2(197355230\sqrt{110}-1144879587) \word{and}\\D&=299795296 \sqrt{110}- 3734763307 +\\
&\qquad\qquad(12 \ln(2\sqrt{110}\m 1)\p 18 \ln(2\sqrt{110}\m 3)-15 \ln(440))(-96294467 \sqrt{110}\p 524292560)\,.
\end{split}
\eeq{AB}
The ratio of the
one loop correction to the tree (classical) result for the potential can be
readily calculated to be $-0.504$. Such ratios are typical of one loop corrections in massless theories (e.g. in massless $\lambda \phi^4$ theory in flat space with a single scalar field the ratio is close to $-1$ \cite{ChengLi}). We discuss in the conclusions how adding gauge field fluctuations can effect this scenario.  

We started with a  classical massless theory, a conformally invariant theory with $\lambda \phi^4$ and $R\,\phi^2$ interactions. After including one loop quantum corrections, the dimensionelss parameter $\lambda$ has been traded for the dimensionful VEV. An important result is that the Ricci scalar of AdS space is now completely determined by the value of the VEV.   

\subsection{Composite operator [E]}

The composite operator [E(y)] is defined via \reff{E}. It appears in the trace but also as a quantum correction to the equations of motion for the scalar fields (see section 4 below). Inserting this operator into an $n$-point Green's function $\Gamma^n(x_1,...,x_n)$ and integrating over all $y$ yields $n$ times the same Green's function \cite{Hathrell}. The Feynman diagrams are therefore identical to those used to evaluate the effective potential $V_{\rm eff}$, namely those of Fig.\ref{RPhi2} (the only difference is that the symmetry factor is multiplied by $n$). The upshot is that $[E]$ can be obtained by taking the negative of the derivative of the one loop part of the effective potential \reff{Veff}, $V_{\rm loop}=V_{\rm eff}-U$, and then multiplying it by $\phi_c$,
\beq
\begin{split}
[E] &= -\phi_c \dfrac{dV_{\rm loop}}{d \phi_c}\\ &= - \dfrac{1}{1152 \pi^2}\Bigg[36\,\Big(\!-\dfrac{R}{3} + \dfrac{\lambda \phi_c^2}{2}\Big) 
  \ln\Big[\,\dfrac{2 R - 3 \lambda \phi_c^2}{2 R}\,\Big]\lambda\,\phi_c^2-108\Big(\!-\dfrac{R}{3} + \dfrac{\lambda \phi_c^2}{2}\Big)^2 \dfrac{\lambda \phi_c^2}{2 R - 3 \lambda \phi_c^2} \\&\quad+ 24\,\Big(\!-\dfrac{R}{3} + \dfrac{\lambda \phi_c^2}{6}\Big) 
  \ln\Big[\,\dfrac{2 R - \lambda \phi_c^2}{2 R}\,\Big]\lambda \phi_c^2 -72 \Big(\!-\dfrac{R}{3} + \dfrac{\lambda \phi_c^2}{6}\Big)^2 \dfrac{\lambda \phi_c^2}{2 R - \lambda \phi_c^2} \\&\qquad+ 10 R \,\lambda\, \phi_c^2 + \lambda^2 \phi_c^4 \Big(18\,\ln\Big[\,\dfrac{2 R}
  {2 R- 3 \lambda M^2}\,\Big]
 +4\ln\Big[\,\dfrac{2 R}
  {2 R- \lambda M^2}\,\Big]\Big)\Bigg]\\ &\qquad\quad -4 \Delta \,.
\end{split}
\eeq{E3} 
  
\section{Equations of motion for the magnetic monopole} 

The quantum-corrected equations of motion for the metric, scalar and
gauge fields are derived in appendix A and are given by equations
\reff{EqMetric2},
\reff{scalar2}, and \reff{EqGauge} respectively. For the magnetic monopole, we seek static spherically symmetric solutions where the spatial symmetry (isometry) and gauge symmetry are both SO(3). These can be viewed as the lowest energy or ground state solution \cite{MB}. The metric, scalar triplet, and non-abelian gauge fields take on the following spherically symmetric form \cite{MB}:
\begin{align}
&\word{metric:} ds^2 = N(r)\,dt^2  - \psi(r) dr^2 -r^2 \,(d\theta^2 +\sin^2(\theta)\,d\phi^2)\label{e1}\\
&\word{scalar:}\phi^a(r) = f(r)\,\dfrac{r^{a}}{r} = f(r)[\sin\theta\sin\varphi,\sin\theta\cos\varphi,\cos\theta]\label{e2}\\
&\word{gauge:} A^{\mu}_a=q(r)\,\xi^{\mu}_a \word{where} \xi^{\mu}_a \word{are the Killing vectors for SO(3), namely}\nonumber\\
\xi^{\mu}_1&=[0,0,\cos\varphi,-\sin\varphi\cot\theta],\xi^{\mu}_2=[0,0,-\sin\varphi,-\cos\varphi\cot\theta] \word{and} \xi^{\mu}_3=[0,0,0,1]\,. 
\label{e3} \end{align}
It will be convenient to work with $a(r)\equiv 1+ r^2 \,q(r)$ instead of $q(r)$. There are four functions of $r$ to determine: the ``metric" fields $N(r)$ and $\psi(r)$, the ``gauge field" $a(r)$ and the ``scalar" field $f(r)$. It is convenient to obtain the equations of motion by direct variation of these functions. The Lagrangian corresponding to $S_{ren}$ is given by
\begin{align}
L&=4\pi\int_{0}^{\infty}\!\!\!\!dr\sqrt{N\,\psi}\,r^2\!\Big(\alpha \,R^2 \p\beta\,R_{\mu\nu}R^{\mu\nu} \m\dfrac{1}{4e^2}\,F_{\mu\nu}^{\,a}\,F^{\mu\nu}_a \p D_{\mu}\phi^a\,D^{\mu}\phi_a \p\dfrac{1}{6}\,R\,\phi^a\phi_a \m \dfrac{\lambda}{4!}(\phi^a\phi_a)^2 \Big)\nonumber\\
&=4\pi\int_0^{\infty}\mathscr{L}\,dr\,.\label{LL}
\end{align}
The quantities that appear in \reff{LL} evaluated using Eqs. \ref{e1}-\ref{e3} are
\begin{align}
R &= \dfrac{-2\psi'}{r\,\psi^2} -\dfrac{N'^2}{2\,\psi\,N^2} +\dfrac{N''}{\psi\,N}-\dfrac{N'\,\psi'}{2\,N\,\psi^2}-\dfrac{2}{r^2}+\dfrac{2}{r^2\,\psi}
+\dfrac{2\,N'}{r\,N\,\psi}\label {RR}\\
R_{\mu\nu}R^{\mu\nu}&= \widetilde{R}_{j}\widetilde{R}^{j}\equiv(\widetilde{R}_{1})^2 +(\widetilde{R}_{2})^2 +(\widetilde{R}_{3})^2 +(\widetilde{R}_{4})^2 \nonumber\\
&\word{where}\qquad \widetilde{R}_{1}\equiv N\,R^{tt}=\dfrac{-\psi'\,N'}{4\,N\psi^2} -\dfrac{N'^2}{4\,\psi\,N^2} +\dfrac{N''}{2\,\psi\,N}
+\dfrac{N'}{r\,N\,\psi}\label{R0}\\
&\qquad\qquad\quad\widetilde{R}_{2}\equiv \psi\,R^{rr}=\dfrac{\psi'}{r\,\psi^2} +\dfrac{N'^2}{4\,N^2\,\psi} -\dfrac{N''\,\psi}{2\,\psi\,N}
+\dfrac{N'\,\psi'}{4\,N\,\psi^2}\label{R1}\\
&\qquad\qquad\quad\widetilde{R}_{3}\equiv r^2\,R^{\theta\theta}=\dfrac{\psi'}{2\,r\,\psi^2} +\dfrac{1}{r^2} -\dfrac{1}{r^2\,\psi}
-\dfrac{N'}{2\,r\,N\,\psi}\label{R2}\\
&\qquad\qquad\quad\widetilde{R}_{4}\equiv r^2\sin^2\theta\,R^{\phi\phi}=\widetilde{R}_{3}\label{R3}\\
F_{\mu\nu}^{\,a}\,F^{\mu\nu}_a =&\dfrac{4\,a'^{\,2}}{r^2\,\psi}+\dfrac{2\,(a^2-1)^2}{r^4} \label{FF}\\
D_{\mu}\phi^a\,D^{\mu}\phi_a=&-\dfrac{f'^{\,2}}{\psi}-\dfrac{2\,a^2\,f^2}{r^2}\label{DD}\\
R\phi^a\phi_a =& \dfrac{f^2}{2\,\psi^2\,r^2\,N^2}\Big\{-4\,r\,\psi'\,N^2-N'^2\,\psi\,r^2+2\,N''\,\psi\,r^2\,N-\psi'\,N'\,r^2\,N-4\,\psi^2\,N^2\nonumber\\
&+4\,\psi\,N^2+4\,N\,N'\,r\,\psi\Big\}\label{RPhi}\\
(\phi^a\,\phi_a)^2 = &f^4 \word{and} \sqrt{-g}=\sqrt{N\,\psi}\,r^2\sin\theta\,.\label{Phi2}\end{align}
Variation with respect to the metric function $N(r)$ yields 
\beq
\dfrac{\partial\mathscr{L}}{\partial N} - \dfrac{\partial}{\partial r} \dfrac{\partial\mathscr{L}}{\partial N'} + \dfrac{\partial^{\,2}}{\partial r^2} \dfrac{\partial\mathscr{L}}{\partial N''}=\dfrac{\sqrt{-g}}{2\sin\theta} \langle T^{tt}\rangle
\nonumber
\eeq{EqN}
The equation of motion for $N$ is
\begin{align}
&\dfrac{\sqrt{N \psi}\,r^2}{2N}(\alpha\,R^2+\beta\widetilde{R}_{j}\widetilde{R}^{j})+ \sqrt{N \psi}\,r^2\Big(2\,\alpha\,R\dfrac{\partial R}{\partial N}+2\,\beta\,\widetilde{R}_{j}\dfrac{\partial \widetilde{R}^{j}}{\partial N}\Big)\nonumber\\&\quad-\Big[\sqrt{N \psi}\,r^2\big(2\,\alpha\,R\dfrac{\partial R}{\partial N'}+2\,\beta\,\widetilde{R}_{j}\dfrac{\partial \widetilde{R}^{j}}{\partial N'}\big)\Big]'+\Big[\sqrt{N \psi}\,r^2\,\big(2\,\alpha\,R\dfrac{\partial R}{\partial N''}+2\,\beta\,\widetilde{R}_{j}\dfrac{\partial \widetilde{R}^{j}}{\partial N''}\big)\Big]''\nonumber\\&\quad\quad\,+\frac{1}{6\,\sqrt{N\,\psi}} \big(f^2\,\m f^2\psi \m f^2 r \dfrac{\psi'}{\psi}\m r^2 f f' \dfrac{\psi'}{\psi}\p 4  rff'\p 2 r^2 f f''\big)-\frac{\lambda\, r^2 f^4 \psi}{48 \sqrt{N \psi}}\nonumber\\&\quad\qquad-\dfrac{\big((a^2-1)^2 \psi+2 r^2 a'^2\big)}{4\,e^2\,r^2 \sqrt{N \psi}}-\frac{a^2 f^2 \psi}{\sqrt{N \psi}}
-\frac{r^2 f'^{\,2}}{6\sqrt{N \psi}}=\dfrac{\sqrt{N\,\psi}\,r^2}{2} \langle T^{tt}\rangle\label{NEq} 
\end{align}
where the Ricci scalar $R$ is given by \reff{RR}, $\widetilde{R}_{j}=\widetilde{R}^{j}$ is given by Eqs.\ref{R0}-\ref{R3} and implicit summation over $j=1,2,3,4$ is assumed.
Variation with respect to the metric function $\psi(r)$ yields the equation 
\beq
\dfrac{\partial\mathscr{L}}{\partial \psi} - \dfrac{\partial}{\partial r} \dfrac{\partial\mathscr{L}}{\partial \psi'} + \dfrac{\partial^{\,2}}{\partial r^2} \dfrac{\partial\mathscr{L}}{\partial \psi''}=-\dfrac{\sqrt{-g}}{2\sin\theta} \langle T^{rr}\rangle 
\eeq{EqPsi}
and we obtain
\begin{align}
&\dfrac{\sqrt{N \psi}\,r^2}{2\psi}(\alpha\,R^2+\beta\widetilde{R}_{j}\widetilde{R}^{j})+ \sqrt{N \psi}\,r^2\Big(2\,\alpha\,R\dfrac{\partial R}{\partial \psi}+2\,\beta\,\widetilde{R}_{j}\dfrac{\partial \widetilde{R}^{j}}{\partial \psi}\Big)\nonumber\\&\,\,-\Big[\sqrt{N \psi}\,r^2\big(2\,\alpha\,R\dfrac{\partial R}{\partial \psi'}+2\,\beta\,\widetilde{R}_{j}\dfrac{\partial \widetilde{R}^{j}}{\partial \psi'}\big)\Big]'+\Big[\sqrt{N \psi}\,r^2\,\big(2\,\alpha\,R\dfrac{\partial R}{\partial \psi''}+2\,\beta\,\widetilde{R}_{j}\dfrac{\partial \widetilde{R}^{j}}{\partial \psi''}\big)\Big]''\nonumber\\
&\quad -\dfrac{\lambda\, r^2 f^4 N}{48\,\sqrt{N\psi}}-\dfrac{N (a^2-1)^2}{4e_R^2\,r^2\,\sqrt{N \psi}}+\dfrac{N^2\,a'^2}{2\,e^2\,(N\psi)^{3/2}}+
\dfrac{N^2 r^2 f'^2}{2\,(N\psi)^{3/2}}\nonumber\\&\quad\quad+\dfrac{r f f' \left(4 N^2+r N'N\right)}{6\,(N\psi)^{3/2}}+ \dfrac{f^2 \left(N^2-N^2\psi-6a^2N^2\psi+rNN'\right)}{6\,(N\psi)^{3/2}}
\nonumber\\&\qquad\qquad= -\dfrac{\sqrt{N\,\psi}\,r^2}{2} \langle T^{rr}\rangle\,. 
\label{PsiEq}\end{align}
Lagrange's equations for the gauge field $a$ is given by
\beq
\dfrac{\partial\mathscr{L}}{\partial a} - \dfrac{\partial}{\partial r} \dfrac{\partial\mathscr{L}}{\partial a'}=0
\eeq{Lagrange_a}
which yields the equation of motion
\beq
2\,a\,(a^2-1) +4\,a\,e^2\,f^2\,r^2 -\dfrac{2}{\psi}\,a''\,r^2 +\dfrac{a'\, r^2}{\psi}\Big(\dfrac{\psi'}{\psi}-\dfrac{N'}{N}\Big)=0\,.
\eeq{Eq_a}
Lagrange's equations for the scalar field $f$ is given by
\beq
\dfrac{\partial\mathscr{L}}{\partial f} - \dfrac{\partial}{\partial r} \dfrac{\partial\mathscr{L}}{\partial f'}= -\dfrac{\sqrt{N\psi}\,r^2}{f}\,[E]
\eeq{Lagrange_f}
which yields the equation of motion
\beq
-\dfrac{4a^2f^2}{r^2} +\dfrac{f^2\,R}{3} - \dfrac{\lambda}{6}\,f^4 + \dfrac{2f\,f''}{\psi} + \dfrac{f\,f'}{\psi}\Big(\dfrac{4}{r} +\dfrac{N'}{N} -\dfrac{\psi'}{\psi}\Big)=-[E]
\eeq{Eq_f}
where $[E]$ is given by \reff{E3} and $R$ is the Ricci scalar given by \reff{RR}.

The above equations of motion are for static field configurations. Therefore the higher derivative terms in the metric field equations \reff{NEq} and \reff{PsiEq} pose no issues as they are spatial not time derivatives. Higher spatial derivatives appear in many branches of physics e.g. in physical acoustics the wave equation is modified by a term with four spatial derivatives when the bending stiffness of a vibrating string is included. 

\subsection{Relation between Ricci scalar of AdS space and the VEV: asymptotic analytical solution}

We now solve the equations of motion  \reff{NEq},\reff{PsiEq}, \reff{Eq_a} and \reff{Eq_f} analytically in the asymptotic region to show that the Ricci scalar of AdS space is determined entirely by the VEV. The vacuum expectation value of the energy momentum tensor in AdS space is given by \reff{TT} and the non-zero components are 
\beq
\begin{split}
&\langle T^{tt}\rangle=\dfrac{1}{4}\dfrac{1}{1+k\,r^2} \langle T^{\mu}_{\,\,\mu}\rangle_0\qquad;\qquad \langle T^{rr}\rangle=- \dfrac{1}{4}(1+k\,r^2) \langle T^{\mu}_{\,\,\mu}\rangle_0\\
&\langle T^{\theta\theta}\rangle=-\dfrac{1}{4\,r^2} \langle T^{\mu}_{\,\,\mu}\rangle_0\qquad\qquad;\qquad\langle T^{\phi\phi}\rangle=- \dfrac{1}{4\,r^2\sin^2\theta}\langle T^{\mu}_{\,\,\mu}\rangle_0
\end{split}
\eeq{EMT2}
where the AdS metric \reff{AdS} was used. The trace $\langle T^{\mu}_{\,\,\mu}\rangle$ is given by
\beq
\begin{split}
\langle T^{\mu}_{\mu}\rangle &= \dfrac{a_2(x)}{16 \pi^2}- [E]\\
&=\dfrac{1}{16 \pi^2}\Big\{\dfrac{1}{60}\Big(R_{\mu\nu\sigma\tau}R^{\mu\nu\sigma\tau} \m R_{\mu\nu}R^{\mu\nu} \p \,\square \,R\Big)
-\dfrac{1}{6}\, F^2 + \dfrac{11}{3} \dfrac{\lambda^2}{4!} \phi^4 - \dfrac{5}{36}\,\lambda \square \,[\phi^2]\} -[E],
\end{split}
\eeq{EnergyM}
where \reff{trace2} and \reff{aa2} were used and $[E]$ is given by \reff{E3}. We now evaluate \reff{EnergyM} in AdS space, the asymptotic spacetime. Asymptotically, $F_{\mu\nu}^a\to 0$, $\square \,[\phi^2] \to 0$, $R_{\rho\sigma\mu\nu}= k\big(g_{\rho\mu}\,g_{\sigma\nu}-g_{\rho\nu}\,g_{\sigma\mu}\big)$ and $R_{\mu\nu}=3\,k\,g_{\mu\nu}$
so that $R=12k$, $\square \,R=0$ and $R_{\mu\nu\sigma\tau}R^{\mu\nu\sigma\tau} \m R_{\mu\nu}R^{\mu\nu}=-12\,k^2$. Substituting these values into \reff{EnergyM}, we obtain
\beq
\langle T^{\mu}_{\mu}\rangle_0=-\dfrac{1}{80}\dfrac{k^2}{\pi^2} +\dfrac{11}{1152}\dfrac{\lambda^2}{\pi^2}\,v^4 -[E]_0
\eeq{TAdS}
where $[E]_0$ is $[E]$ evaluated in asymptotic AdS space.

The boundary conditions for the magnetic monopole \cite{MB} are that asymptotically, as $r\to \infty$, the spacetime is AdS where $N\to1+ k r^2$, $\psi\to1/(1+k \,r^2)$, $f\to v$, $r\,f' \to 0$ ($f'$ drops off faster than $1/r$), $r^2\,f'' \to 0$, $a\to 0$ and $a'\to 0$. Both $v$ and $k$ are positive constants. Substituting these boundary conditions into the gravity equation \reff{NEq} (or \reff{PsiEq}) and using \reff{TAdS} yields the following relation 
\beq
\dfrac{v^2}{2} -\dfrac{\lambda v^4}{48 \,k}=\dfrac{1}{8\,k}\Big(-\dfrac{1}{80}\dfrac{k^2}{\pi^2} +\dfrac{11}{1152}\dfrac{\lambda^2}{\pi^2}\,v^4 -[E]_0\Big)\,.
\eeq{GravAdS}
We can eliminate $[E]_0$ above by solving the scalar equation \reff{Lagrange_f} asymptotically. This yields
\beq
4\,k\,v^2-\dfrac{\lambda v^4}{6}=-[E]_0
\eeq{ScalarAdS}
where we used that $R=R_0=12\,k$ in AdS space. Substituting the above into \reff{GravAdS} yields the solution
\beq
k= \dfrac{\sqrt{110}}{12} \lambda \,v^2
\eeq{kk}
so that
\beq
R_0=\sqrt{110}\,\lambda\,v^2 \,.
\eeq{RAdS}
The Ricci scalar of AdS space is therefore determined solely by the VEV since the value of $\lambda$ is known (it is no longer a free parameter having been traded for the dimensionful VEV). Substituting \reff{kk} into \reff{ScalarAdS}, yields $[E]_0=-\frac{\lambda\,v^4}{6}\,(\,2\sqrt{110}-1)$. This agrees with expression \reff{E3} asymptotically i.e. after substituting $R=\sqrt{110}\,\lambda\,v^2$ and $\phi_c=v=M$.

The remaining unbroken U(1) is associated with $F^{\mu\nu}_3$ and the magnetic field is defined via $ F^{ij}_3=\epsilon^{ijk} B^k$. Asymptotically, 
$a(r)\to 0$ and the function $q(r)$ appearing in \reff{e3} approaches $-1/r^2$. It is easy to verify that one obtains a radial magnetic field that varies as $1/r^2$ at large distances, corresponding to a magnetic monopole. 





\section{Conclusions}

In previous work \cite{MB}, a magnetic monopole solution in AdS space
was obtained without introducing explicitly a mass term. In that calculation, spontaneous symmetry breaking (SSB)
of gauge symnmetry responsible for the magnetic monopole occurred via gravitation
itself through the coupling term $R \phi^2$ in a conformally invariant
action. This works as long as a length scale is introduced by hand because classically there is no length scale. In the present work, we introduced a renormalization scale into the massless theory by considering quantum corrections. Symmetry breaking was radiatively induced $\grave{a}$ la Coleman-Weinberg \cite{Coleman}, albeit in
a more complicated massless theory containing gravity where the one
loop effective potential must take into account the $R \phi^2$
interaction in addition to the usual $\lambda
\phi^4$. The dimensionless $\lambda$, defined at the renormalizaton scale $M$, was traded for the
dimensionful VEV and the Ricci scalar of the background AdS spacetime
was determined entirely by the VEV. Though we discussed the quantum
corrections of a classical conformal invariant theory in the context
of the magnetic monopole, the techniques and results presented here
could potentially have wider consequences. For example, in a recent article \cite{Bars}, scale and conformal symmetry are presented
as fundamental principles for physics and cosmology. The authors have
a model containing a Higgs field $H$, a dilaton field $\phi$, standard
model fields as well as gravity. The authors point out that in a
conformally invariant theory there is no mechanism at the classical
level to set the scale of $\phi_0$, the minimum of the dilaton field
$\phi$. They then mention that quantum corrections may alleviate this
problem in a fashion that is reminiscent to \cite{Coleman}; our calculation
provides a concrete implementation of this proposal.

Our work can now be naturally extended in a few ways. First, one can add
gauge field fluctuations in the calculation of the effective potential. Then the
coupling contant $\lambda$ can be expressed in terms the electromagnetic coupling constant
$e$ as in \cite{Coleman}. In that scenario, the two parameters in the theory become $e$ and the dimensionful VEV. Second, we used the symmetry of AdS space to solve for $\langle T_{\mu\nu} \rangle_0$, the asymptotic value of $\langle T_{\mu\nu} \rangle$. This allowed us to solve the equations of motion analytically in the asymptotic regime and to obtain an expression relating the Ricci scalar of AdS space to the VEV. The interior spacetime obeys spherical symmetry but not the symmetry of AdS space. Therefore, the finite and nonlocal part of $\langle T_{\mu\nu} \rangle$ for the interior would require a more elaborate calculation. One could then obtain numerical solutions of the interior. It is of interest to see how these numerical solutions containing quantum corrections in an AdS background compare with those obtained in the classical context of General Relativity (GR)\cite{Numeric}. Third, we worked with a background AdS spacetime because the Ricci scalar of AdS space had the right sign for
classical SSB \cite{MB}. However, the VEV here is obtained from the quantum-corrected effective potential. In the presence of quantum corrections, de Sitter (dS) space could well be a viable background spacetime. The Ricci scalar of the background AdS space was determined solely by the VEV, which should apply to dS space as well. Such solutions could be of
greater cosmological interest.

\section*{Acknowledgments}
AE acknowledges support from an NSERC discovery grant. He thanks KITP
for a four week stay during the summer of 2012 where part of this work
was completed. This research was, supported in part by the National
Science Foundation under Grant No. NSF PHY11-25915. NG was supported in part by the National Science Foundation (NSF)
through grant PHY-1213456.

\begin{appendix}
\def\theequation{A.\arabic{equation}}
\setcounter{equation}{0}

\section{Equations of motion with quantum corrections}

The equations of motion for the metric, non-abelian gauge
fields, and scalar triplet are obtained by variation of the
total action $S=S_{\rm ren} \p W_{\rm ren}$ with respect to each field. For the metric we obtain 
\beq
\dfrac{2}{\sqrt{-g}}\dfrac{\delta}{\delta g^{\mu\nu}} S_{\rm ren} = - \langle T_{\mu\nu}\rangle
\eeq{EqMetric1}
where we used \reff{TT}. With $S_{\rm ren}$ given by \reff{action3} the metric field equations are 
\begin{align}\label{EqMetric2}
&\qquad\qquad\qquad\alpha \,H_{\mu\nu} + \beta \,K_{\mu\nu} + M_{\mu\nu}=-\langle T_{\mu\nu}\rangle\quad \\\word{where}\nonumber\\
&H_{\mu\nu}\equiv\dfrac{2}{\sqrt{-g}}\dfrac{\delta}{\delta g^{\mu\nu}}\int R^2\, \sqrt{-g}\,d^{\,4}x=4\nabla_{\mu}\nabla_{\nu}R-4\,g_{\mu\nu}\square R-g_{\mu\nu}R^2+4R\,R_{\mu\nu}\nonumber\\
&K_{\mu\nu}\equiv\!\dfrac{2}{\sqrt{-g}}\dfrac{\delta}{\delta g^{\mu\nu}}\!\!\int\! \!R_{\alpha\beta}R^{\alpha\beta} \sqrt{-g}\,d^{\,4}x\nonumber \\&\qquad\qquad\eq 4\nabla_{\nu}\nabla_{\alpha}R_{\mu}^{\,\,\alpha}\m 2\,\square R_{\mu\nu}-g_{\mu\nu}\square R \p 4R_{\mu}^{\,\,\alpha}R_{\alpha\nu}\m g_{\mu\nu}R_{\alpha\beta}R^{\alpha\beta}.\nonumber\\
&M_{\mu\nu}\!\equiv\!\dfrac{2}{\sqrt{-g}}\dfrac{\delta}{\delta g^{\mu\nu}}\!\!\int\! \!\Big\{-\dfrac{1}{4e_R^2}\,F^2+ (D\phi)^2 +\dfrac{1}{6}\,R\,\phi^2 - \lambda_R^2\,\phi^4 \Big\} \sqrt{-g}\,d^{\,4}x\nonumber\\
&\qquad =\dfrac{1}{4e_R^2}g_{\mu\nu}F^2 -\dfrac{1}{e_R^2}\,F_{\mu\beta}^{\,a}\,{F_{\nu}^{\,\beta}}_a +2\,D_{\mu}\phi^a\,D_{\nu}\phi_a -g_{\mu\nu}\big((D\phi)^2- \lambda_R^2\,\phi^4\big)\nonumber\\
&\qquad+\dfrac{1}{3}\big(g_{\mu\nu}\nabla_{\alpha}\nabla^{\alpha}\phi^2- \nabla_{\mu}\nabla_{\nu}\phi^2 +(R_{\mu\nu}-\dfrac{1}{2}\,g_{\mu\nu}R)\phi^2\big)\nonumber\,.
\end{align}
For the scalar field, we have $\tfrac{2}{\sqrt{-g}}\tfrac{\delta}{\delta \phi^a}(S_{\rm ren}+W_{\rm ren}) =0$, which yields the equation 
\beq
\Big(2\,D^{\mu}D_{\mu}\phi_a- \big(\dfrac{R}{3}-4\lambda_R^2\phi^2\big)\phi_a\Big)\phi^a =\dfrac{1}{\sqrt{-g}}\phi^a \dfrac{\delta W_{\rm ren}}{\delta \phi^a}=[E].      
\eeq{scalar2} 
  For the gauge field we simply have $\tfrac{2}{\sqrt{-g}}\tfrac{\delta}{\delta A_{\nu}}S_{\rm ren} =0$, since $\tfrac{2}{\sqrt{-g}}\tfrac{\delta }{\delta A_{\nu}}W_{\rm ren}$ is zero when evaluated in the vacuum state where $F_{\mu\nu}^a \to 0$ and $A_{\mu}^a$ is a pure gauge which we set to zero. The equation of motion for the gauge field is then 
\begin{align}\label{EqGauge}
\nabla_{\mu}F^{\mu\nu}_{\,a}+\varepsilon_{a}^{\,\,bc}A_{{\mu}_b}\,F^{\mu\nu}_{\,c}=2\,e_R \,\varepsilon_{a}^{\,bc}\,D^v\!\phi_{b}\,\phi_c \,.
\end{align}
\end{appendix}

\end{document}